\documentclass[useAMS,usenatbib,usegraphicx,a4paper]{mn2e}

\title[High precision microlensing maps]
  {High precision microlensing maps of the Galactic bulge}
\author[Kerins, Robin \& Marshall]
   {E.~Kerins$^1$, A.C.~Robin$^2$ and D.J.~Marshall$^{2,3}$\\
    $^1$Jodrell Bank Centre for Astrophysics, The University of Manchester,
Oxford Road, Manchester M13 9PL, UK\\
    $^2$Observatoire de \bes\, Institut UTINAM, Universit\'e
Franche-Comt\'e, CNRS-UMR 6213, BP 1615, 25010 \bes\, Cedex, France\\
   $^3$D\'epartement de physique, de g\'enie physique et
d'optique, Universit\'e Laval, Qu\'ebec, QC, G1K 7P4, Canada}

\newcommand{\be}{\begin{equation}}
\newcommand{\ee}{\end{equation}}
\newcommand{\bes}{Besan\c{c}on}

\newcommand{\tres}{\tau_{\rmn{res}}}
\newcommand{\tdia}{\tau_{\rmn{dia}}}
\newcommand{\tsc}{\tau_{\rmn{sc}}}

\begin{document}

\maketitle

\begin{abstract}
We present detailed maps of the microlensing optical depth and event density
over an area
of 195~deg$^2$ towards the Galactic bulge. The maps are computed from
synthetic stellar catalogues generated from the \bes\ Galaxy Model, which
comprises four stellar populations and a three-dimensional extinction map
calibrated against the Two-Micron All-Sky Survey. The optical depth maps have a
resolution of 15~arcminutes, corresponding to the angular resolution of the
extinction map.
We
compute optical depth and event density maps for all resolved sources
above $I=19$, for unresolved  (difference image) sources magnified above this
limit, and
for bright standard candle sources in the bulge. We show that the resulting
optical
depth contours are dominated by extinction effects, exhibiting fine
structure in stark contrast to previous theoretical optical depth maps.
Optical depth comparisons between Galactic models and optical microlensing
survey
measurements cannot safely ignore extinction or assume it to be
smooth.
We
show how the event distribution for hypothetical $J$ and $K$-band
microlensing
surveys, using existing ground-based facilities such as
VISTA, UKIRT or CFHT, would be much less affected by extinction, especially in
the
$K$ band. The near infrared provides a substantial
sensitivity increase over current $I$-band surveys and a more faithful
tracer of the underlying stellar distribution, something which upcoming
variability surveys such as VVV will be able to exploit. Synthetic
population models offer a promising way forward to fully exploit large
microlensing datasets for Galactic structure studies.
\end{abstract}


\section{Introduction}

Several microlensing survey teams have been monitoring millions of stars over a
large region of the Galactic bulge for more than a decade \citep{moa,macho,ogle,eros}.
These surveys have detected thousands of
events and the combined detection rate of the  OGLE-III and
MOA-II surveys is currently around 800 unique events per year. The
microlensing optical depth, that is the instantaneous number of ongoing
microlensing events per source star, is a key measurable for these surveys and
provides an important constraint on the bulge surface mass
density. Its
dependence upon direction provides, in principle, a powerful probe
of
the three-dimensional geometry of the bulge stellar mass distribution.

To date several measurements have been made of the optical depth along
different directions. The MOA survey \citep{moa} find an optical
depth from difference image analysis (DIA) of 28 events of $\tdia =
2.59^{+0.84}_{-0.64} \times 10^{-6}$ for bulge and disc sources
around Baade's Window $(l \simeq 1\degr, b \simeq -3\fdg9)$. Most recent
measurements of
the optical depth have been based upon a restricted subset of red clump giant
(RCG) stars, which act approximately as standard candles and are known to be confined to the bulge. From a sample of 42
RCG events the MACHO collaboration \citep{macho} measures a standard candle
optical depth of
$\tsc = 2.17^{+0.47}_{-0.38}\times 10^{-6}$ at $(l = 1\fdg 5, b =
-2\fdg 7)$. The OGLE-II survey \citep{ogle} reports a value of
$\tsc = 2.55^{+0.57}_{-0.46} \times 10^{-6}$ determined from 32 RCG events
in the direction $(l = 1\fdg 2, b = -2\fdg 7)$. The largest single sample
analysed to date involves 120 RCG events obtained by the EROS-2 experiment
\citep{eros}, from which it measured the optical depth over a large region,
including $\tsc = (2.42\pm 0.47) \times 10^{-6}$ within $(-3\fdg 5 < b <
-1\fdg 4)$.  EROS-2 also reports an optical depth gradient with
latitude $|b|$ of $d\tsc/d|b| \simeq (0.78\pm 0.27) \times
10^{-6}$~deg$^{-1}$ at $b = -2\fdg 7$, in line with similar gradient
determinations by MACHO and OGLE-II.

Despite initial disagreement, optical depth measurements from RCG sources are
 now largely in accord with recent theoretical models
\citep{eva02,han03,wood05} and support the existence of a bar-like bulge
which is oriented $10-20\degr$ from the Sun---Galactic Centre line. However
RCG events represent less than 10\% of all bulge microlensing
events. A major limitation to the study of the remaining events
comes from the current simplicity of theoretical models. A fully
developed model must account for the line of sight distribution of the
sources in both the disc and bulge components. This distribution is a
function not only of the underlying density models but also of the
experimental sensitivity, the source luminosity function and the distribution
of dust. The latter factor is particularly problematic for studies of the
spatial distribution of events since extinction varies strongly with sky
position. Until recently realistic three dimensional extinction maps were
unavailable.

In this paper we present the first microlensing optical depth maps to
incorporate a realistic three-dimensional extinction map. The map was
developed by \cite{mar06} as part of the \bes\ Galaxy Model \citep{rob03}.
In Section~\ref{mod} we briefly overview the \bes\ Model including
how the three-dimensional dust map was generated. In Section~\ref{odep} we
describe how we set up the
optical depth calculations. In Section~\ref{maps} we present maps of optical
depth and of event density for current $I$-band microlensing surveys, as well
as for a hypothetical near-infrared survey which, as we  show,  would be
much less affected by
dust. We end with a discussion in Section~\ref{disc}.

\section{The Besan\c{c}on Galaxy Model} \label{mod}

The \bes\ Galaxy Model is a simulation tool aimed at testing
galaxy evolution scenarios by comparing stellar distributions predicted
by these scenarios with observational constraints, such as photometric
star counts and kinematics. A complete description of the model
ingredients can be found in \cite{rob03}. The model assumes that the Sun is at a
Galactocentric distance $R_0 = 8.5$~kpc. We
summarise here the model's principal features which are relevant to the
present analysis. In particular, we do not discuss the details of the
kinematics and metallicity of each population as the microlensing
optical depth does not depend on them.

\subsection{Stellar populations}

The model assumes that stars are created from gas following a star
formation history and an initial mass function; stellar evolution
follows evolutionary tracks. To reproduce the overall galaxy formation
and evolution we distinguish four populations of different ages and
star formation history.

\begin{itemize}
\item The spheroid population is described in \cite{Robin2000} as a single
burst population of age 14 Gyr. Its density in the central regions is small
compared to other
populations and it is therefore only a marginal contributor to the optical
depth.
\item The thick disc is formed from stars born
about 11-12 Gyr ago in a short period of time as implied by recent
metallicity determinations for this population. We also assume a single
burst for simplicity.  For the thick disc, star formation occurred from
the gas already settled in the disc.  Kinematics, deduced from
observational constraints  \citep{Ojha96,Ojha99}, imply that it has
undergone a merging event shortly
after the disc formation  \citep{Robin96}, increasing the
disc thickness and giving a higher velocity dispersion and scale
height.  The initial mass function (IMF), density distribution and local
normalisation are
constrained from star counts \citep{Reyle2001}.
\item The bulge population is present in the centre of the Galaxy and
extends to about 3-4 kpc. Its age is around 10$\pm 2$ Gyr.
The population is modeled from \cite{Bruzual2003} population synthesis, as a
single burst and a Salpeter IMF.
 This population reflects a triaxial bar
distribution, as determined by
\cite{Picaud2004} from near-infrared star counts. The angle
of the major axis of the bulge with the sun-Galactic Centre axis
is found to be $11\fdg 3$. The scale lengths are 1.590~kpc, 0.424~kpc
and 0.424~kpc. The total bulge mass is $2.05 \times
10^{10}~\mbox{M}_{\sun}$.
\item For the thin disc a standard
evolutionary model is used to compute the
disc population, comprising an IMF, a star formation rate
(SFR) and a set of evolutionary tracks, as described in
\cite{Haywood97b} and references therein.  The disc population is assumed to
evolve during
10 Gyr, with multiple periods of star formation
creating seven distinct age populations.  A set of IMF slopes and SFRs are
tentatively assumed and then
tested against star counts. The tuning of disc parameters against
relevant observational data has been described in
\shortcite{Haywood97a,Haywood97b}.
A revised IMF has been adjusted to agree with the most recent Hipparcos
results:
the
age-velocity dispersion relation is from \cite{Gomez97}, the local
luminosity function from \cite{Jahreiss97}, giving an IMF
slope $\alpha$ = 1.5 in the low mass range (0.5-0.08~M$_{\sun}$), in good
agreement with
\cite{Kroupa2001}.  The scale height has been computed
self-consistently using the potential via the Boltzmann equation. The
local dynamical mass is taken from \cite{Creze1998}.
\end{itemize}

Stellar remnants are also included in the model, though their role in the
optical depth analysis is only indirect in contributing to the overall
mass of lensing stars and in limiting the overall mass normalisation for
visible stars which act as microlens sources; the microlensing optical depth
is insensitive to the stellar or remnant mass function of the lenses (see
Section~\ref{odep}).
We assume all white dwarfs to be of type  DA and use evolutionary tracks and
atmosphere models from \cite{Bergeron01}, complemented by
\cite{Chabrier99} for the very cool end, applicable to the halo.
No brown
dwarfs are included in the present model. All simulated stars are single
stars.

The evolutionary model fixes the distribution of stars in the space of
intrinsic parameters: effective temperature, gravity, absolute
magnitude, mass and age. These parameters are converted into colours in
various systems through stellar atmosphere models.

The detectability of stars computed in the simulations is subject to
interstellar extinction
and
to observational errors. The latter are computed assuming Gaussian
distributions
with a dispersion as a function of the star magnitude. The extinction
distribution is obtained as described below.

\subsection{Extinction distribution}

\cite{mar06} have built a 3D model of the distribution of
the dust in the inner Galaxy ($|l|<100\degr, |b|<10\degr$) from an analysis of
the 2MASS survey \citep{Skrutskie2006}
using the \bes\ Galaxy Model.  By comparing the observed reddened
stars to the unredenned simluated stars of the \bes\ model, they
were able
to calculate the extinction as a function of distance for any given
line of sight in the Galaxy. The extinction along a line of sight is
that which minimises the chi-squared difference between the $J-K_s$
colour distributions of the observed and simulated stars. The final
map has an angular resolution of 15~arcmins and a distance resolution of
100-500 pc. Better angular resolutions are possible but at the cost
of reducing the resolution in distance.

Having established the Galactic model we now turn our attention to its
microlensing properties.

\section{Optical depth evaluation} \label{odep}

The optical depth is one of the primary measurables which can be obtained
from statistical analyses of microlensing events. It depends only on the
distribution and mass normalisation of the source and lens stellar
populations and is therefore potentially a very powerful probe of Galactic
structure. The optical depth to a source at distance $S$ is \citep{pac86}
   \be
       \tau = \int_0^S \frac{\pi R_{\rm E}^2}{M} \rho(L) \, dL,
\label{tau}
   \ee
where $L$ is the observer--lens distance, $M$ is
the lens mass, $\rho$ is the lens mass density and $R_{\rm E} = [4 G M
L(S-L)/c^2 S]^{1/2}$ is the Einstein radius of the lens, with $G$ and $c$ having
their usual meaning.

An evaluation of $\tau$ must involve an average over the observed source
population. Such an average must take account of, among other factors, the
survey sensitivity. Current microlensing surveys such as OGLE-III and MOA-II are
observing in the $I$ band and are capable of $4\%$ photometry at approximately
$I = 19$
\citep[c.f. Figure~5 of][]{uda02}. We therefore adopt this limit for our
$I$-band calculations. We shall also consider a hypothetical
near-infrared microlensing survey observing in the $J$ or $K$ band. We
assume it to be capable of $4\%$ photometry at $J = 18$ and $K = 17$ for
comparison.
Such a survey could be
conducted with facilities such as UKIRT, VISTA or CFHT and, as we shall see,
would
have the advantage of being much less affected by dust than the $I$-band
surveys. One survey which will be capable of detecting bulge microlensing
events in the near-infrared is the Vista Variables in the Via Lactea (VVV)
survey, which is scheduled to commence soon after the completion of VISTA in
2009.

We start by computing optical depths according to equation~(\ref{tau}) over a
three dimensional grid of source positions $(l,b,S)$, where $l$ and $b$ are
Galactic coordinates. For the lenses we employ the same underlying mass
density distributions
as are used in the \bes\ Model simulations. We then use the \bes\ population
synthesis program
to generate synthetic catalogues of millions of source stars. The catalogues
cover the
region $|l| < 9\fdg 75$, $|b| < 5\degr$, giving a total
area of
195~deg$^2$. Each catalogue probes 3239 lines of sight to give a map
resolution
of $15\arcmin \times 15 \arcmin$, which is the limit of the resolution of the
dust map. Finally, we interpolate the optical depth grid at the position of each
synthetic source to arrive at a source-averaged optical depth, $\tau$, for each
line of sight.

We consider three different optical depth quantities. The first is $\tres$,
which
is the optical depth averaged over all ``resolved'' sources, i.e. those which
at
baseline are brighter than the assumed survey limit of $I = 19$, $J = 18$ or
$K = 17$. The
second optical depth quantity is $\tdia$ which represents the average over all
``difference image analysis''  (DIA) sources, i.e. those which at peak
magnification are brighter than the assumed survey limit. DIA sources can be
significantly fainter than the survey limit at baseline and so our
simulated source catalogues
go four magnitudes fainter
than
our assumed survey limits.  We weight the optical depth contribution of each DIA
source by $\mbox{min}(1,u_{\rm t}^2)$, where $u_{\rm t}$ is the largest impact
parameter, in units of $R_{\rm E}$, which allows the event to be detected above
the survey limit. This is necessary because, at any instant, only a fraction
$u_{\rm t}^2$
of unresolved sources are expected to be magnified sufficiently to be
detectable as events. Many recent optical depth measurements have been based
upon a
subset of events involving bright red clump giant (RCG) sources, which are
approximate standard candles. We therefore compute a third optical depth
quantity, $\tsc$, where $\tau$ is averaged over a subset of bright
``standard candle'' sources. Our standard candle source selection has
   \be
      I_0 < \min [I^*_{\rm 0,RC}+1,9(V-I)_0 + I^*_{\rm 0,RC} -5.5],
\label{rcgcut}
  \ee
where subscript 0 denotes de-reddened
magnitude and the red clump magnitude $I^*_{\rm 0,RC} = 14.66-0.033l$, with
$l$ the Galactic longitude expressed in degrees. This prescription
follows the ``extended red clump'' selection algorithm of \cite{ogle}, where
we have adopted a weighted linear fit to the zero-point corrected red clump
magnitude values, $I^*_{\rm 0,RC}$, listed
in Table~2 of \cite{ogle} for Galactic latitudes $b < 0$.

\section{Optical depth and Event Density maps} \label{maps}

\begin{figure*}
\includegraphics[scale=0.5,angle=270]{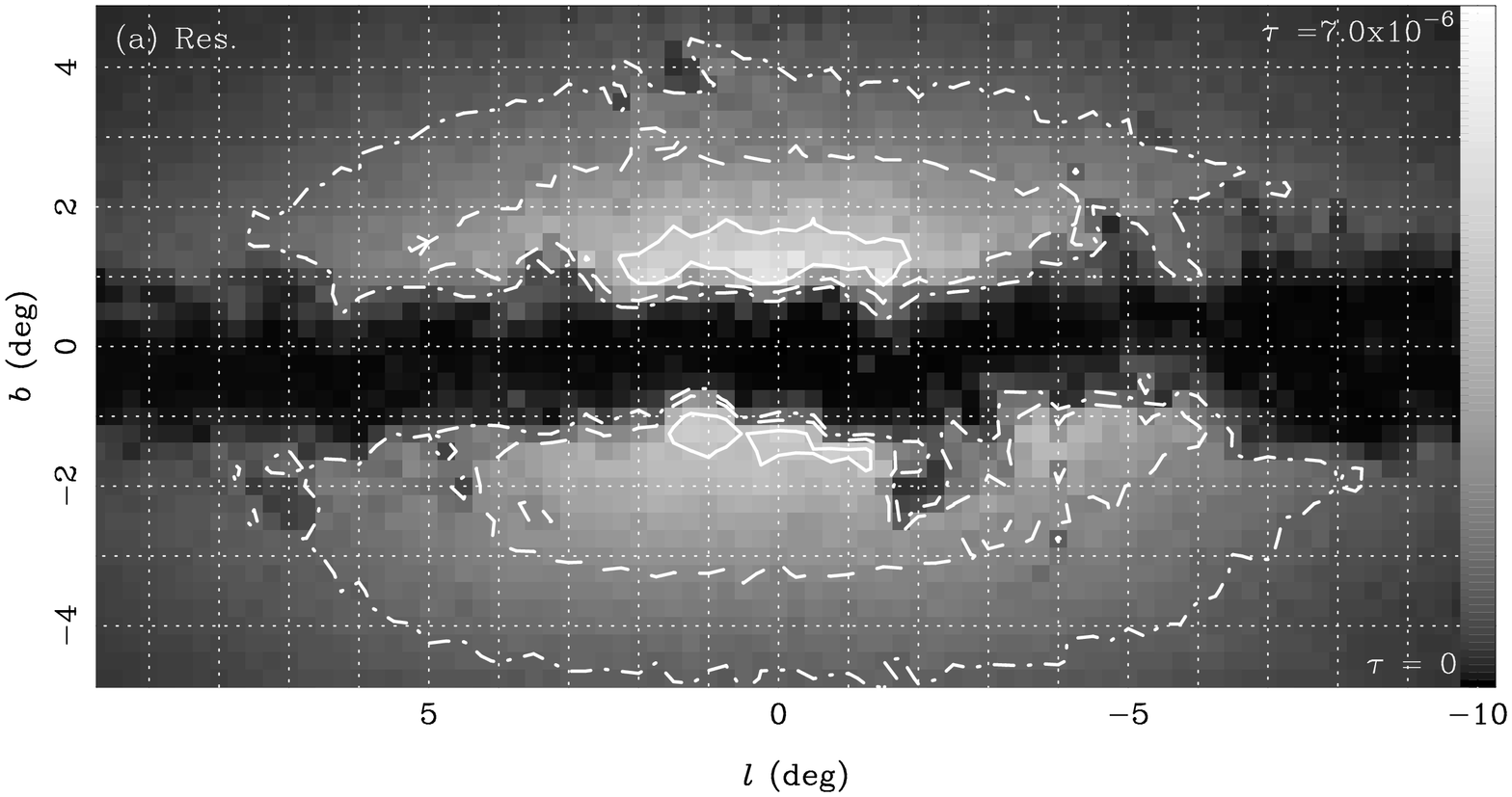}
\includegraphics[scale=0.5,angle=270]{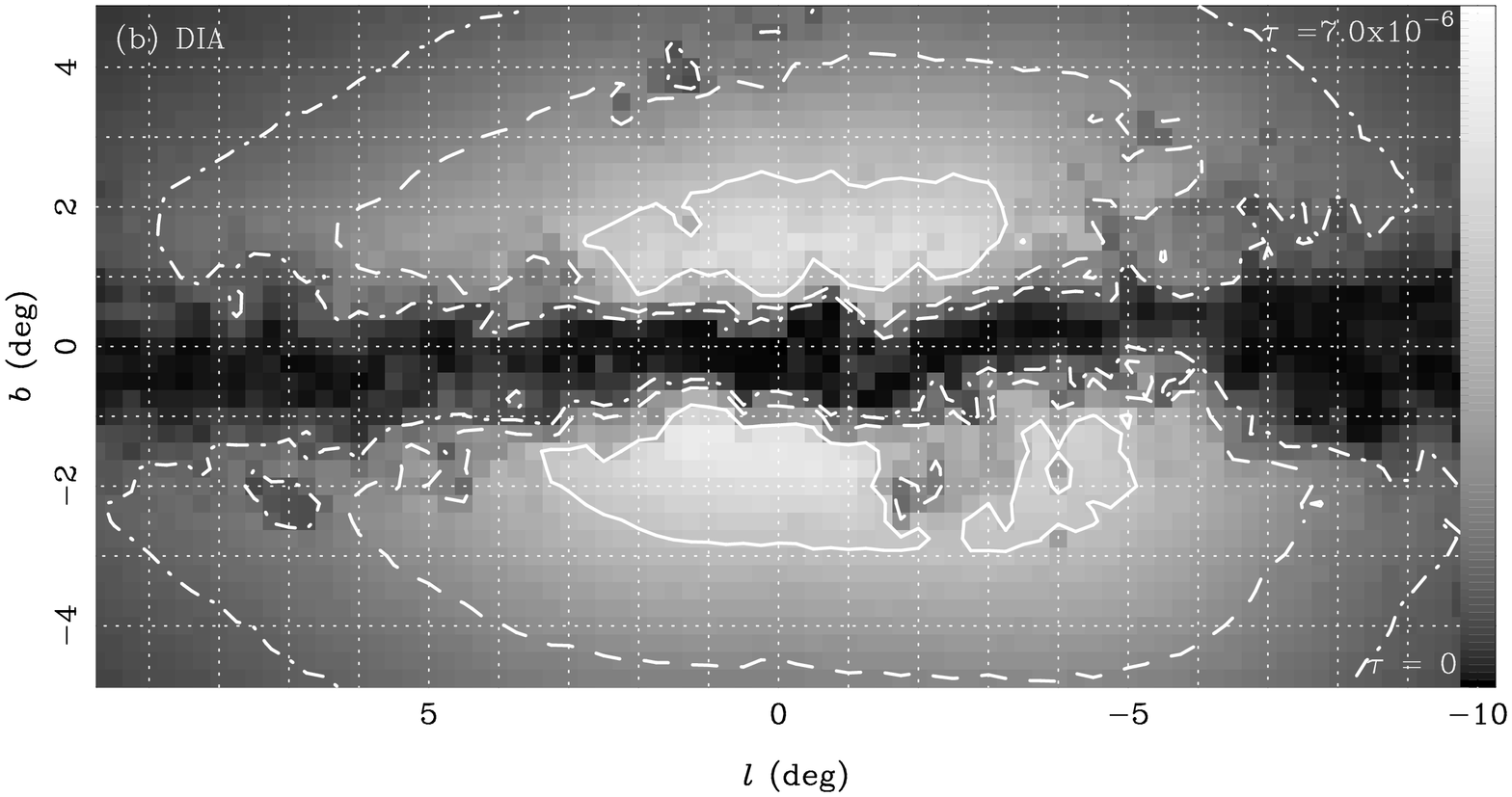}
\includegraphics[scale=0.5,angle=270]{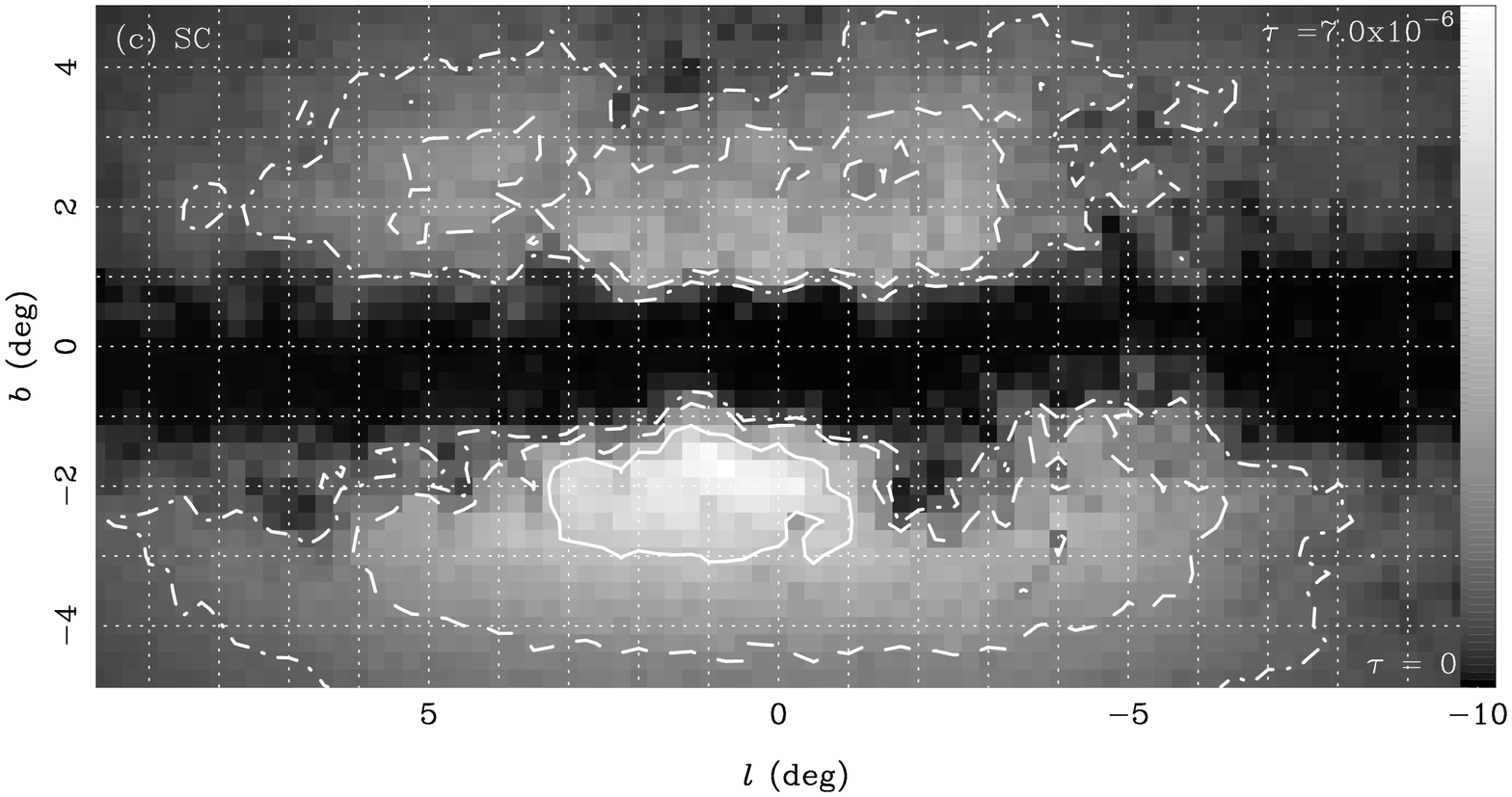}
\caption{$I$-band optical depth greyscale maps. (a) The
optical depth, $\tres$, to all sources brighter then $I = 19$ at baseline. (b)
The optical depth, $\tdia$, to all sources brighter than $I = 19$ at peak
magnification. (c) The optical depth, $\tsc$, to ``standard candle'' sources
as
defined by equation~(\ref{rcgcut}). Solid, dashed and dot-dashed contours
indicate optical depths of 4, 2 and $1\times 10^{-6}$, respectively. The low
optical depth towards the Galactic plane is due to the high
extinction in this region. The greyscale range is given by the bar to the
right of each plot and is normalised to a maximum value of $7 \times
10^{-6}$ for all panels.}
\label{odepmap}
\end{figure*}

\begin{figure*}
\includegraphics[scale=0.5,angle=270]{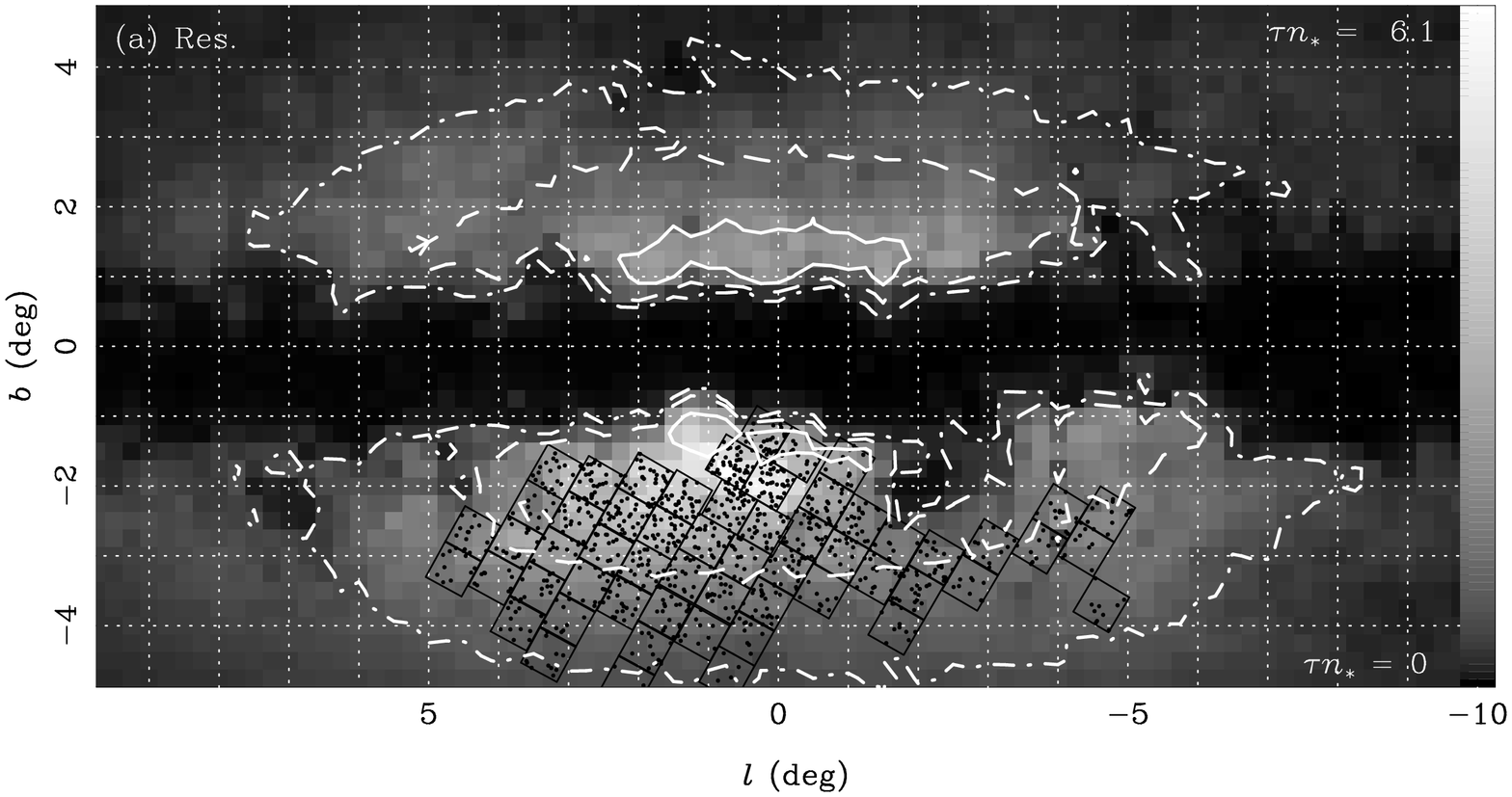}
\includegraphics[scale=0.5,angle=270]{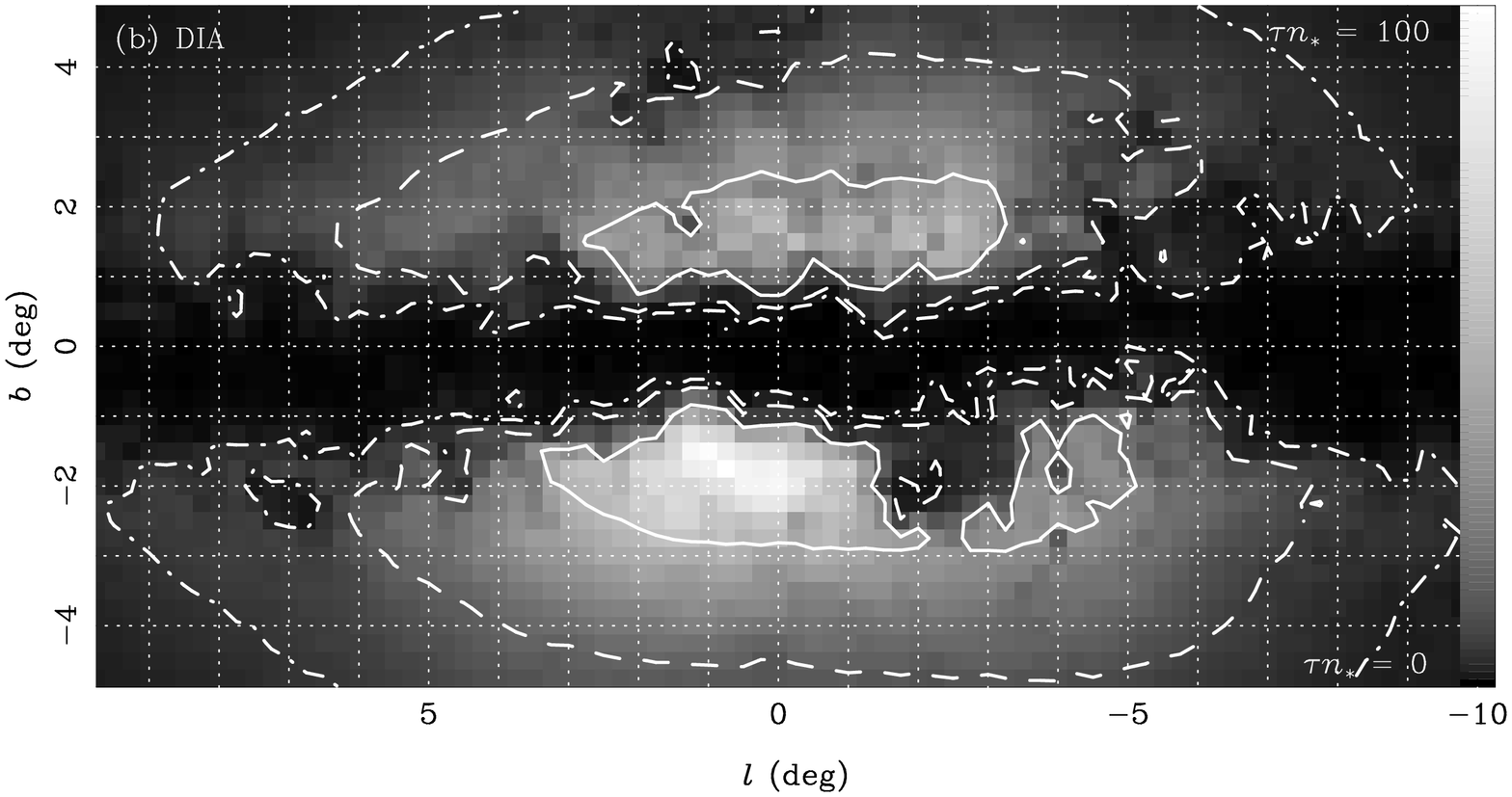}
\includegraphics[scale=0.5,angle=270]{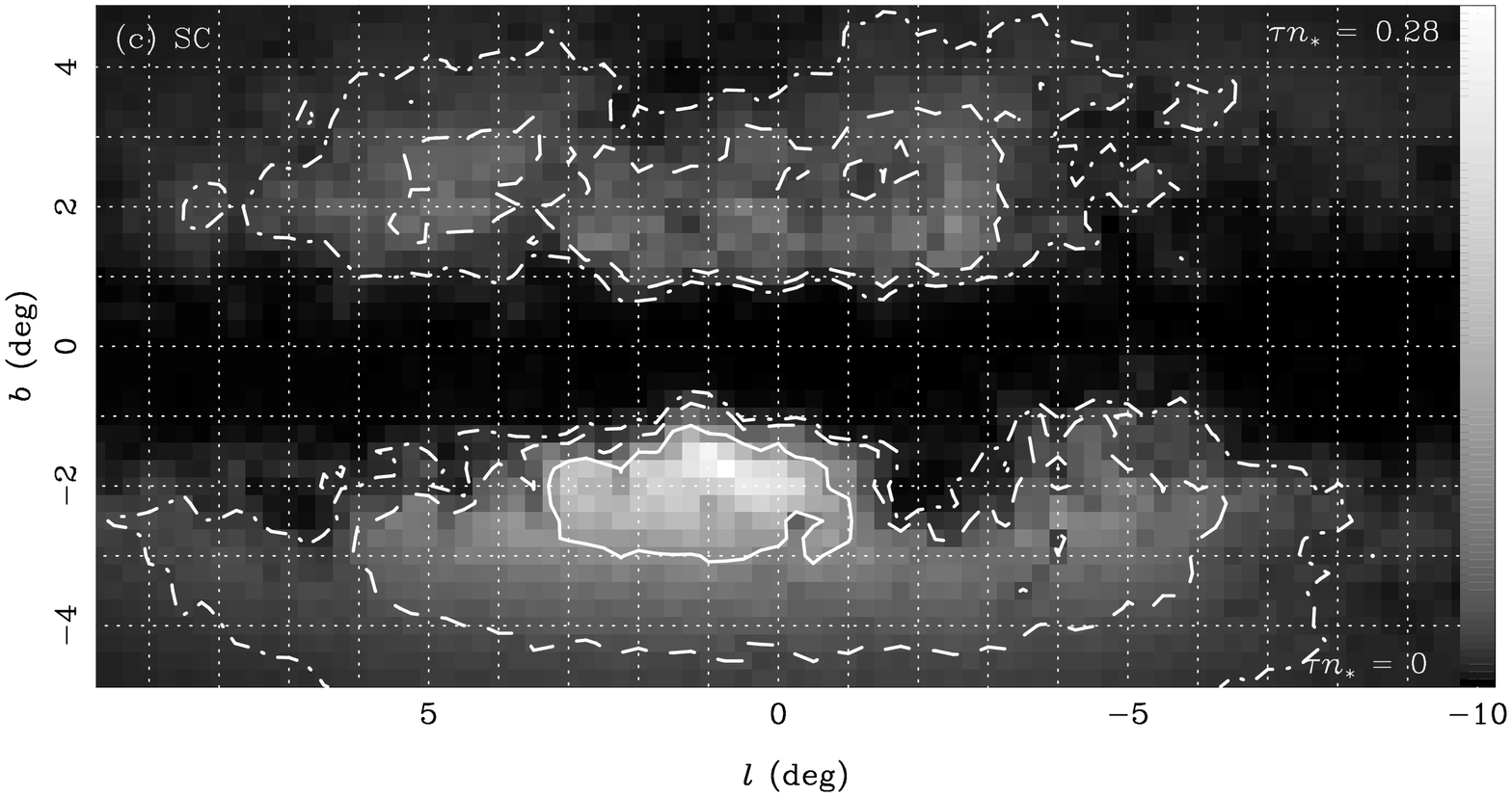}
\caption{$I$-band grayscale maps of instantaneous event density, that is
the
product of
optical depth and source number density $(\tau n_*)$. The units for $\tau n_*$
are events/deg$^2$. Panels (a), (b) and (c) correspond to the cases plotted in
Figure~\ref{odepmap}. For comparison the optical depth contours in
Figure~\ref{odepmap} are also overplotted here in white. The top panel also
shows
57 OGLE-III field locations (black squares) along with 1069 events
with Einstein radius crossing times $10 < t_{\rm E} / \mbox{days} < 100$
detected by the OGLE Early Warning System between 2005 and 2007 (black
dots). Note that the grayscale normalisation is different in each of the
plots.}
\label{evden}
\end{figure*}

We present the results from Section~\ref{odep} in two forms. Firstly as
straightforward $I$-band maps of optical depth and secondly we present $I
$-band maps of event
density, which we define as the product of optical depth and source number
density. The latter provides an instantaneous snapshot of the distribution
of microlensing events on the sky. In the last part of this section we also
present near-infrared microlensing maps.

\subsection{Maps of Optical Depth} \label{taumap}

In Figure~\ref{odepmap} we present $I$-band optical
depth maps for resolved sources ($\tres$), DIA sources ($\tdia$) and bulge
``standard candle'' sources ($\tsc$). The optical depth is indicated both by the
greyscale level and by the contours, which are set at 1, 2 and $4\times
10^{-6}$.

The most obvious distinguishing feature of these maps from previously
published
predictions is the irregularity of the contour line shapes. This is a direct
consequence of the effect of dust. At latitudes $|b| \la 1\degr$ the very high
column density of dust results in a negligible optical depth because only
relatively nearby sources are detectable and these have an intrinsically small
lensing optical depth from equation~(\ref{tau}). The region of high extinction
centred at $l \simeq -2\degr, b \simeq -2\degr$ is also clearly evident from
the
microlensing
optical depth distribution. Also noticeable is the steep optical depth
gradient towards the plane at $|b| \simeq 1\degr$. Away
from the
Galactic plane the optical depth contours are smoother and comparable to
previous models \cite[e.g.][]{wood05}. In particular the predicted ``standard
candle'' optical depth gradient south of the plane and away from regions of
high dust density is $d\tau_{\rmn sc}/d|b| \simeq 0.7 \times
10^{-6}$~deg$^{-1}$, in very good agreement
with measurements from RCG sources \citep{macho,ogle,eros}.

In the absence
of the dust the central hotspot in the optical depth and event density maps
would be located at negative Galactic longitude due to the higher optical
depth towards the far side of the bar \citep[e.g.][]{wood05}. However, this
is much more difficult to see in the presence of the dust distribution.
When the dust is included the $I$-band optical depth peaks in a localised
hotspot centred around
$l \simeq 0\fdg5, b \simeq -2\degr$ with a value of $7\times 10^{-6}$ for
standard candle sources in the bulge, $5.7\times 10^{-6}$ for resolved
sources and $6
\times 10^{-6}$ for DIA sources. Evidence for such a hotspot comes from
EROS-2 \citep{eros} and from the OGLE-II Early Warning System
\citep[EWS,][]{uda03}. The
EROS-2 bright star event sample clearly shows the greatest contribution from
the
measured optical depth coming from events in this area \citep[c.f.
Figure~14 in][]{eros}. In Section~\ref{evmap} we discuss the distribution of
events from
the OGLE-III EWS.

One interesting feature which emerges from Figure~\ref{odepmap} is that the
DIA optical depth $\tdia$ appears less prone to the effects of
extinction, with a noticeably narrower avoidance zone due to dust near the
Galactic plane. Also interesting are differences in the optical depth
between the three measures for specific locations. For example, at Baade's
Window $(l = 1\degr, b = -3\fdg9)$, both $\tdia$ and $\tsc$ yield optical
depths $\tau \simeq 2.5\times 10^{-6}$ whilst $\tres \simeq 1.5 \times
10^{-6}$.

Comparison of the optical depth maps with recent
optical depth measurements \citep{macho,ogle,eros} indicates a tendency of
the model to overestimate the optical depth of bulge standard candle sources
towards the plane at (at $b \ga
-3\degr$). One factor which may be contributing to this is the relatively
large Galactocentric distance scale employed in the model ($R_0 = 8.5$~kpc).
Equally, this tendency may also indicate that the bulge model and resulting
extinction map
may not be optimal. Another contributing factor may be that the extinction map
effectively saturates in higher extinction regions $(A_V > 20)$ where the
2MASS $K$-band sensitivity is relatively low. Consequently, the
extinction may be underestimated in regions where $A_V > 15$ (Schultheis et
al, in preparation).
Either way, a new bulge model for the \bes\ simulation is in
preparation and microlensing data should provide a useful
consistency check for it.

Another important factor when comparing predicted and observed optical
depths is the
photometric cuts used to define the source sample from which the
optical depth is computed. The resolved source optical depth $\tres$
produces a significantly lower optical depth than either of the other
two measures, highlighting the need for theoretical optical
depth calculations to consider the photometric characteristics of
individual survey selections. Whilst survey detection efficiency calculations
may account for factors such as colour selections, they nonetheless assume
some underlying luminosity function for the microlensing sources, which may
not necessarily be the same as assumed in theoretical calculations.

\subsection{Maps of Event Density} \label{evmap}

Figure~\ref{evden} shows equivalent $I$-band maps for the instantaneous event
number density, which we define as the product of source-averaged optical depth
and the source number density. This quantity is indicative of the spatial
variation in the event rate provided there are not significant spatial gradients
in the average event duration. The grayscale map in
the plots in Figure~\ref{evden} trace the event density and, to aid
comparison, we have overplotted the optical depth
contours from Figure~\ref{odepmap}. In the upper panel of Figure~\ref{evden} we
plot the positions of 57 OGLE-III fields together
with 1069 events detected within these fields between 2005 and 2007 by the
OGLE-III
EWS\footnote{\tt http://www.astrouw.edu.pl/$\sim$ogle/ogle3/ews/ews.html}. We
have excluded events with
Einstein radius crossing times $t_{\rm E} < 10$~days in order to suppress the
bias induced by the fact that some of the OGLE-III fields are monitored
several times per night whilst others are monitored nightly. Additionally we
exclude long-duration events with $t_{\rm E} > 100$~days so as to minimise
the risk of tracing kinematic structures that would not be represented
in our maps which are based only on optical depth.

For all the $I$-band maps there is a predicted hot-spot in the event density at
around $l \simeq 1\degr, b \simeq -2\degr$.The maps clearly indicate that one
expects many
more events
south of the plane than north of it,
supporting the current strategy of the survey teams in focussing predominately
on the southern side of the plane.

Interestingly the distribution of OGLE-III EWS alerted events also
shows hotspots in the regions predicted by the model, and generally the model
correlates quite well with the alert event distribution. Of course the
detection efficiency of the EWS events is completely uncalibrated so one
must be wary of drawing too much from such a comparison, but on the face of it
the model appears capable of reproducing clumpy features seen in the observed
event distribution. One exception may be at $l \simeq 1\fdg 5, b \simeq
-1\fdg 5$ where
the model predicts a strong $I$-band microlensing signature for resolved
sources but which is not covered by the OGLE survey, which has been optimised
over 16 seasons in order to concentrate on high event yield locations.
Nonetheless, with sufficiently sophisticated models there is clearly
the exciting prospect of using the large available samples of microlensing
events to constrain fine structure in the
underlying Galactic structure. However, to do this the detection efficiency of
the
event samples must be known.

\subsection{Near-infrared microlensing}

We finish by considering the potential impact of microlensing surveys in the
near-infrared \citep{gou95,eva02}. Specifically, we consider the event
density in the $J$ and $K$-bands for resolved
sources,
assuming a survey capable of $4\%$ photometry at $J = 18$ and $K = 17$. Such
a survey could
be performed using the wide-field infrared cameras available on telescopes
such as UKIRT, VISTA or CFHT. Indeed the Vista Variables in the Via Lactea
(VVV) survey is scheduled to commence on VISTA in 2009 as an ESO Public
Survey. VVV will survey some 520~deg$^2$ of the disk and bulge, primarily
in the $K$ band, over five seasons, with at least one season of almost
nightly monitoring from which a substantial number of microlensing events
should be detected.

\begin{figure}
\includegraphics[scale=0.35,angle=270]{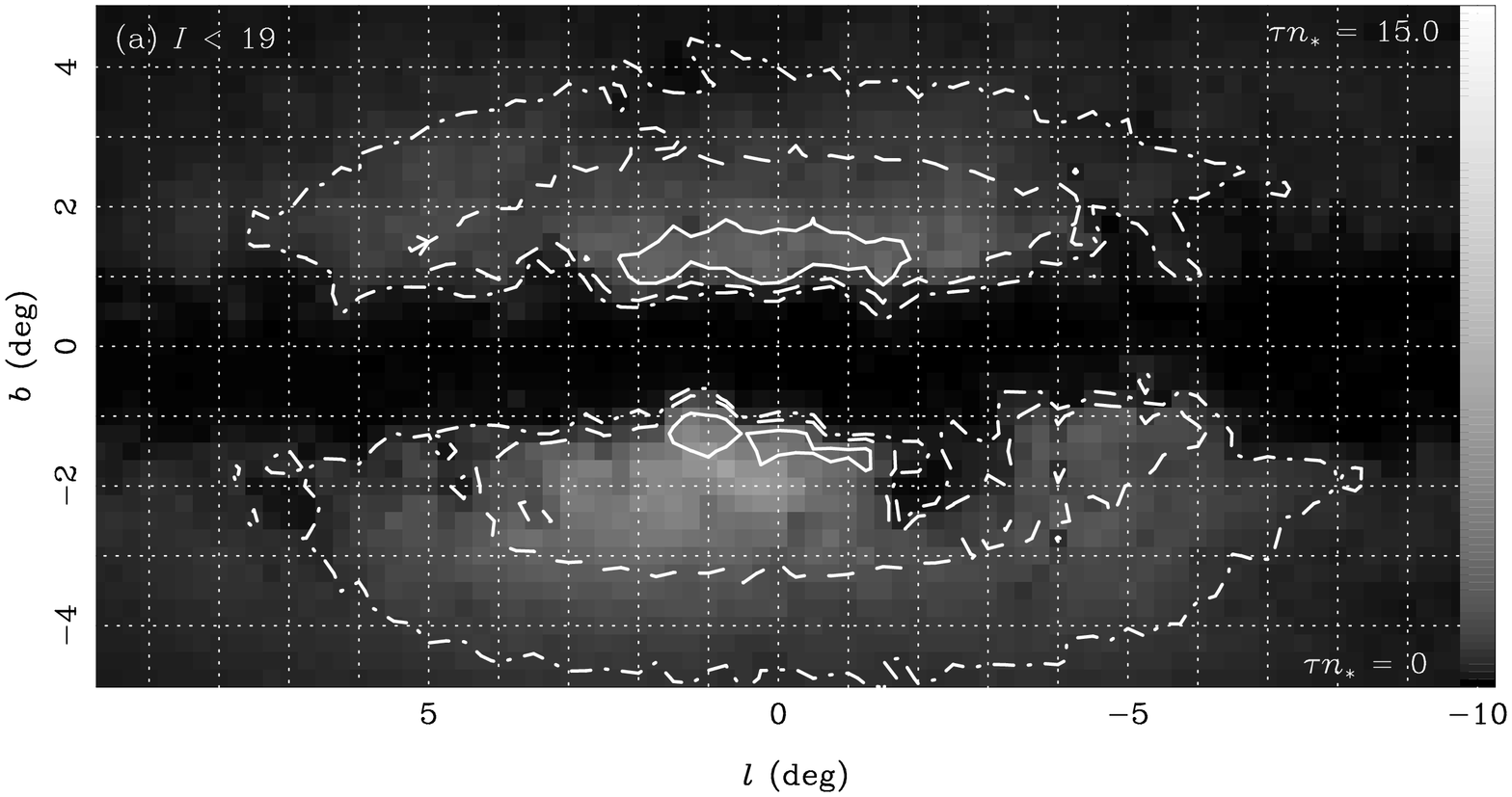}
\includegraphics[scale=0.35,angle=270]{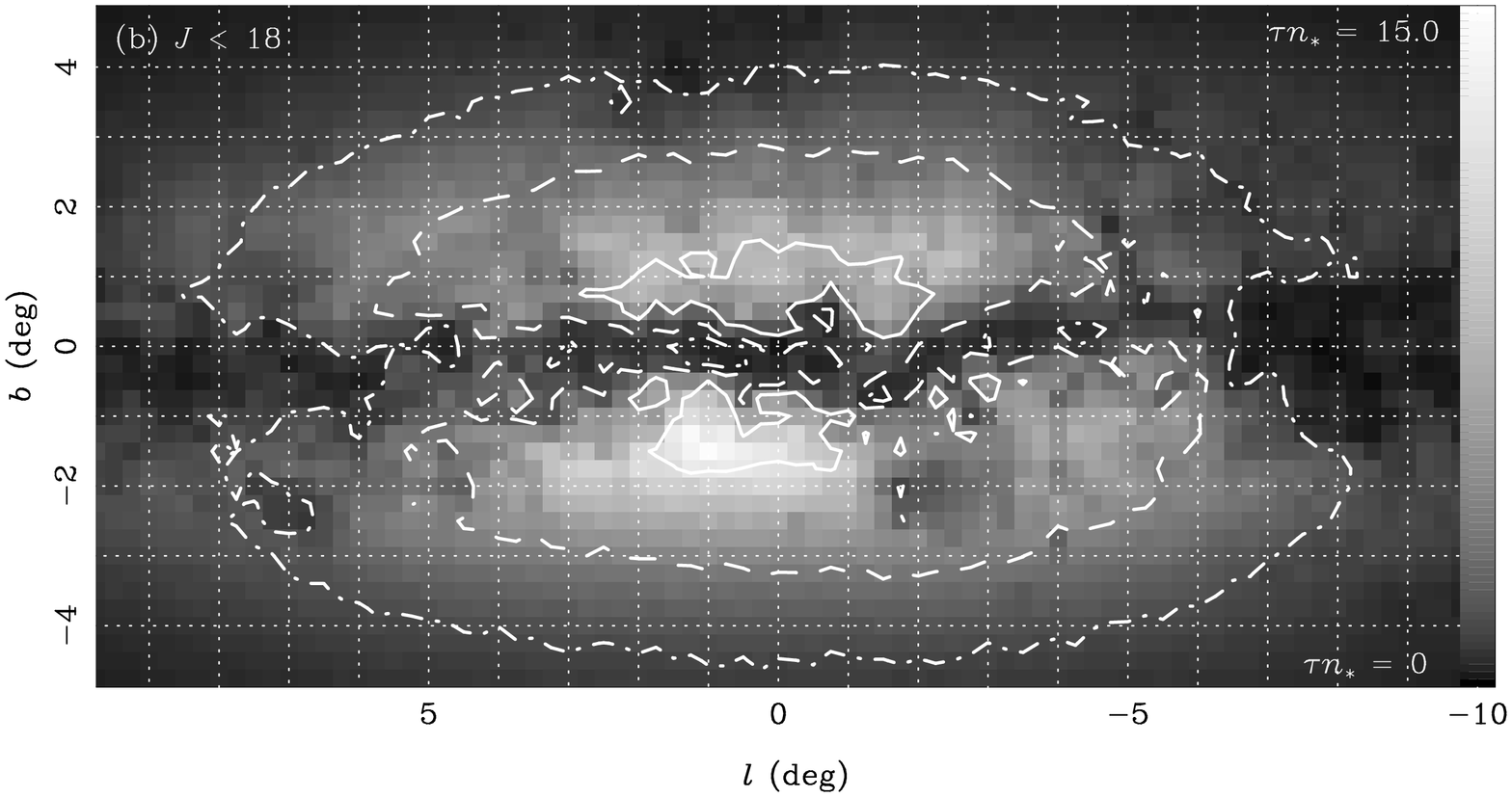}
\includegraphics[scale=0.35,angle=270]{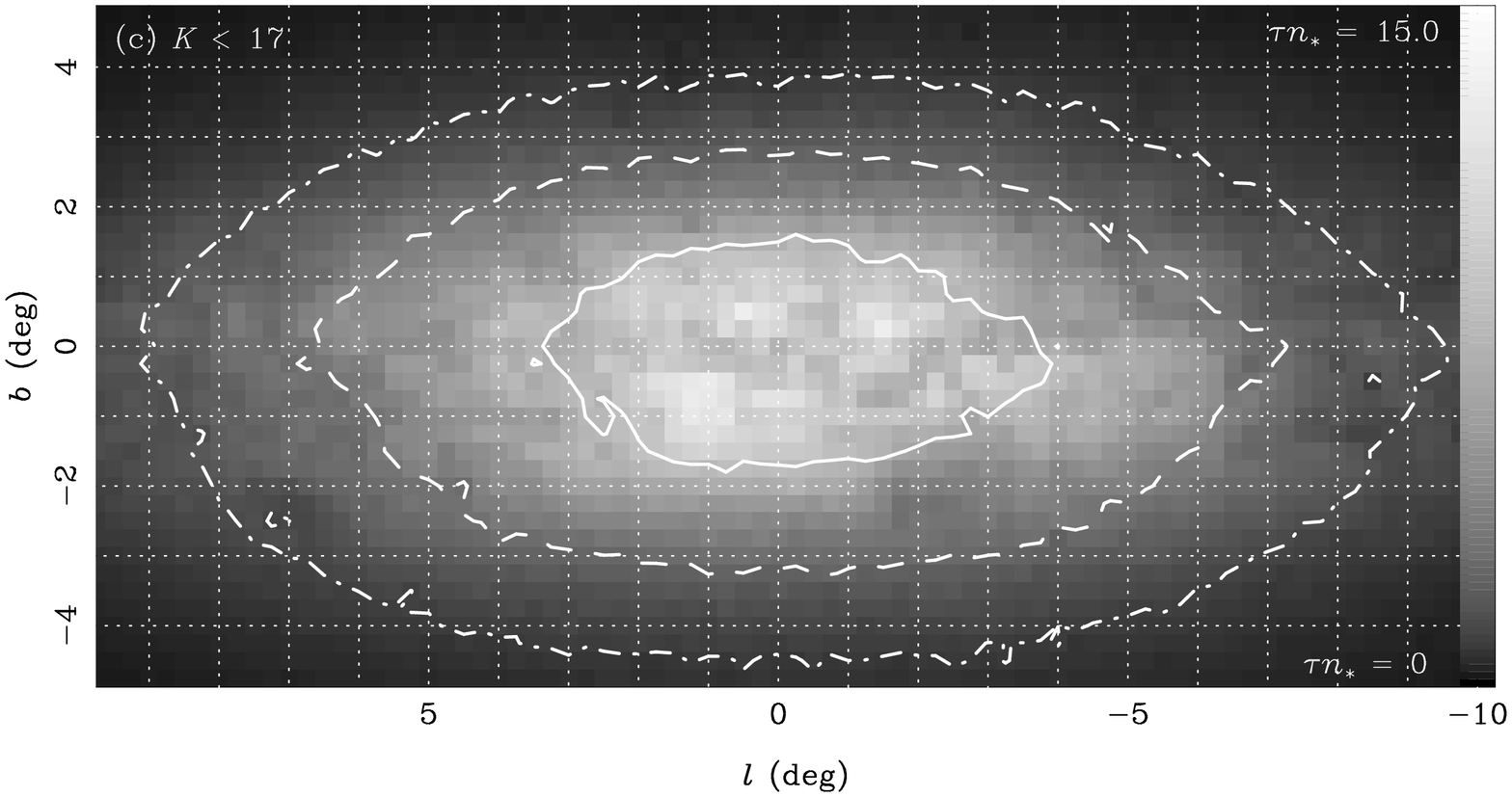}
\caption{A comparison of $I$, $J$ and $K$-band maps of instantaneous event
density.
The grayscale in each map is normalised to a peak density of 15
events/deg$^2$.
Panel (a) is a reproduction of panel~(a) in Figure~\ref{evden} whilst
panels~(b) and (c)
shows the event density in the $J$ and $K$ bands, respectively, for all
sources brighter than $J = 18$ and $K =
17$. Such a near-infrared microlensing survey could be undertaken with
UKIRT, VISTA or CFHT. The advantage of
lesser extinction, particularly in the $K$ band, is clearly evident. The
contours indicate optical depth (not event density),
with levels set the same as for Figure~\ref{odepmap}.}
\label{kband}
\end{figure}

The
top, middle and bottom panels in Figure~\ref{kband} show,
respectively, the $I$-band map (also shown in the top panel of
Figure~\ref{evden}), the $J$-band map and the $K$-band map of event density
for resolved sources (i.e. using $\tres$). The
grayscale levels of the maps are matched for direct comparison and, as in
Figure~\ref{evden}, the
contours indicate the optical depth. The advantage of
observing
in the near-infrared is very clear. Firstly the effects of extinction are
greatly reduced, especially in $K$, so
that the event density is much higher and, as attested by the smoothness of the
$K$-band optical depth contours, more faithfully traces the shape of the
underlying mass
distribution. These two factors alone mean that a $K$-band survey like
VVV would have
powerful advantages over $I$-band surveys for Galactic structure studies.
However the other main advantage of observing in $J$ or $K$ is that the
northern
hemisphere also becomes an attractive target for microlensing studies. In
particular
the region around $|l| \la 2\degr, b \simeq 1\fdg5$ exhibits a strong
microlensing
hotspot in the $K$ band, and a reasonably strong hotspot in $J$. The
significance of this is that this region includes the
far side of the bar at negative $l$. The ability to survey both the near
and far sides of the
bar would allow the separate microlensing contributions of the disc and bulge
to
be separated, as the far bar should have a significantly higher disc optical
depth than the near bar.

\section{Discussion} \label{disc}

We have used a synthetic population synthesis model of the Galaxy to
construct maps of the microlensing optical depth and event density. The
\bes\ Model employs a three-dimensional extinction map
and allows optical depth maps to be computed for individual passbands, and
for the maps to adopt sophisticated cuts in source colour and magnitude.

We have computed $I$-band optical depths maps for three distinct event
samples: stars which at baseline are resolved at $I < 19$ ($\tres$),
comparable to current microlensing survey limits, stars which reach $I < 19$
at peak magnification ($\tdia$), and sources which are selected using colour
cuts similar to those used to define standard candle samples such as red
clump giants ($\tsc$). The structure in the resulting optical depth
maps is dominated by the effects of extinction towards the Galactic pane, and
demonstrates a much stronger microlensing signal South of the Galactic
plane, which is where most of the survey effort is being concentrated. We
find that $\tdia$ tends
to be less affected by the dust. More generally, the comparison between the
observed and modelled optical depth can be quite sensitive to the
photometric characteristics of the event selection.

We also compute maps of event density, that is the product of optical depth
and source density. We find good correlation between these maps and the
location of 1069 events detected by the OGLE-III Early Warning System (EWS).
A direct quantitative comparison cannot be made since the efficiency of the
EWS is unknown. However the model successfully predicts the region of
highest event yield as seen by both the OGLE-III and EROS-2 surveys,
suggesting that the large scale features of the model are correct. We also
find good agreement between the predicted optical depth gradient with
latitude and that determined by microlensing surveys. However, the
model presented here tends to over-predict the bulge optical depth towards
the plane in the case of standard candle source samples. This may possibly
be due to a combination of a relatively large assumed Galactocentric distance
and an underestimation of extinction in high extinction regions. An improved
Galactic model is currently being developed.

We also investigate the potential of near-infrared microlensing surveys by
computing event density maps in the $J$ and $K$ bands. The $K$ band in
particular is extremely promising in circumventing the worst effects of
interstellar dust. $K$-band optical depth contours more faithfully trace the
underlying
structure of the disc and bulge and also allow both the near- and far-bar to
be probed for microlensing. Comparison of near- and far-bar microlensing
samples would allow the separate microlensing signatures of disc and bulge
lenses to be decoupled, providing additional constraints on Galactic models.
Upcoming near-infrared variability surveys, such as VVV, have the potential
to provide excellent microlensing samples for Galactic structure studies.

The number of discovered microlensing events is approaching 5000, and yet
so far only a few hundred have been analysed for Galactic structure studies.
Whilst theoretical models which ignore extinction (or assume it to be smooth)
can be compared reasonably successfully with samples of $\sim 100$ events,
they likely will be inadequate for dealing with
samples of several thousand events, where fine structure may start to become
evident. An accurate calibration of
the stellar mass budget in the inner Galaxy is also a vital
prerequisite to evaluating the allowable mass budget
of dark matter in the inner Galaxy. The analysis of microlensing datasets
can therefore provide important indirect constraints on the dark matter
halo profile.
We believe that synthetic modelling provides an excellent way forward for
such work. By fully exploiting the large and expanding catalogues of
microlensing samples, synthetic modelling provides the potential
for microlensing
to be used as a high-precision probe of Galactic structure.

\section*{Acknowledgements}
EK is supported by an Advanced Fellowship from the  Science and Technology
Facilities Council. DJM is supported by the Natural Sciences and Engineering
Research
Council of Canada through its SRO programme. EK wishes to thank Shude Mao for
helpful
discussions. EK and ACR would like to thank
James Binney for hosting a
European Science Foundation Workshop on Galaxy modelling, from which
came the idea for this work.


\begin{thebibliography}{}

\bibitem[\protect\citeauthoryear{Bergbush \& VandenBerg}{1992}]{bv92}
Bergbush P., VandenBerg D., 1992, ApJS, 81, 163

\bibitem[\protect\citeauthoryear{Bergeron, Leggett,
\& Ruiz}{2001}]{Bergeron01} Bergeron P., Leggett S.~K., Ruiz M.~T.,
2001, ApJS, 133, 413

\bibitem[\protect\citeauthoryear{Chabrier}{1999}]{Chabrier99}
Chabrier G., 1999, ApJ, 513, L103

\bibitem[\protect\citeauthoryear{Bruzual
\& Charlot}{2003}]{Bruzual2003} Bruzual G., Charlot S., 2003, MNRAS,
344, 1000

\bibitem[\protect\citeauthoryear{Cr\'ez\'e et
al.}{1998}]{Creze1998} Cr\'ez\'e M., Chereul E., Bienaym\'e O., Pichon
C., 1998, A\&A, 329, 920

\bibitem[\protect\citeauthoryear{Dorman}{1992}]{dor92}
Dorman B., 1992, ApJS, 81, 221

\bibitem[\protect\citeauthoryear{Dwek et al.}{1995}]{dwek}
Dwek E. et al., 1995, ApJ, 445, 716

\bibitem[\protect\citeauthoryear{Evans \& Belokurov}{2002}]{eva02}
Evans N.W., Belokurov V., 2002, ApJ, 567, 119

\bibitem[\protect\citeauthoryear{Gomez et
al.}{1997}]{Gomez97}
Gomez A.~E., Grenier S., Udry S., Haywood M., Meillon L., Sabas V., Sellier
A., Morin D., 1997, ESASP, 402, 621

\bibitem[\protect\citeauthoryear{Gould}{1995}]{gou95}
Gould A., 1995, ApJ, 446, L71

\bibitem[\protect\citeauthoryear{Hamadache et
al.}{2006}]{eros} Hamadache C., et al., 2006, A\&A, 454, 185

\bibitem[\protect\citeauthoryear{Han \& Gould}{2003}]{han03}
Han C., Gould A., 2003, ApJ, 592, 172

\bibitem[\protect\citeauthoryear{Haywood, Robin,
\& Cr\'ez\'e}{1997a}]{Haywood97a} Haywood M., Robin A.~C., Cr\'ez\'e M.,
1997a, A\&A, 320, 440

\bibitem[\protect\citeauthoryear{Haywood, Robin,
\& Cr\'ez\'e}{1997b}]{Haywood97b} Haywood M., Robin A.~C., Cr\'ez\'e M.,
1997b, A\&A, 320, 428

\bibitem[\protect\citeauthoryear{Jahrei{\ss}
\& Wielen}{1997}]{Jahreiss97} Jahrei{\ss} H., Wielen R., 1997,
ESASP, 402, 675

\bibitem[\protect\citeauthoryear{Kroupa}{2001}]{Kroupa2001} Kroupa
P., 2001, MNRAS, 322, 231

\bibitem[\protect\citeauthoryear{Lejeune, Cuisinier \& Buser}{1997}]{lej97}
Lejeune T., Cuisinier F., Buser R., 1997, A\&AS, 125, 229

\bibitem[\protect\citeauthoryear{Lejeune, Cuisinier \& Buser}{1998}]{lej98}
Lejeune T., Cuisinier F., Buser R., 1998, A\&AS, 130, 65

\bibitem[\protect\citeauthoryear{Marshall et al.}{2006}]{mar06}
Marshall D.J., Robin A.C., Reyl\'e C., Schultheis M., Picaud S., 2006,
A\&A, 453, 635

\bibitem[\protect\citeauthoryear{Ojha et al.}{1996}]{Ojha96}
Ojha D. K., Bienaym\'e O., Robin A. C., Cr\'ez\'e M., Mohan V., 1996, A\&A,
311, 456

\bibitem[\protect\citeauthoryear{Ojha, Bienaym\'e, Mohan \&
Robin}{1999}]{Ojha99}
Ojha D. K., Bienaym\'e O., Mohan, V., Robin A. C., 1999, A\&A, 351, 945

\bibitem[\protect\citeauthoryear{Paczy\'nski}{1986}]{pac86}
Paczy\'nski B., 1986, ApJ, 304, 1

\bibitem[\protect\citeauthoryear{Picaud
\& Robin}{2004}]{Picaud2004} Picaud S., Robin A.~C., 2004, A\&A,
428, 891

\bibitem[\protect\citeauthoryear{Popowski et al.}{2005}]{macho}
Popowski P. et al., 2005, ApJ, 631, 879

\bibitem[\protect\citeauthoryear{Reyl{\'e}
\& Robin}{2001}]{Reyle2001} Reyl{\'e} C., Robin A.~C., 2001, A\&A,
373, 886

\bibitem[\protect\citeauthoryear{Robin et
al.}{1996}]{Robin96} Robin A.~C., Haywood M., Cr\'ez\'e M., Ojha
D.~K., Bienaym\'e O., 1996, A\&A, 305, 125


\bibitem[\protect\citeauthoryear{Robin, Reyl\'e \&
Cr\'ez\'e}{2000}]{Robin2000}
Robin A., Reyl\'e C., Cr\'ez\'e M., 2000 A\&A, 359, 103

\bibitem[\protect\citeauthoryear{Robin, Reyl\'e, Derri\`ere \&
Picaud}{2003}]{rob03}
Robin A., Reyl\'e C., Derri\`ere S., Picaud S., 2003, A\&A, 409, 523

\bibitem[\protect\citeauthoryear{Skrutskie et
al.}{2006}]{Skrutskie2006} Skrutskie M.~F., et al., 2006, AJ, 131,
1163

\bibitem[\protect\citeauthoryear{Sumi et al.}{2003}]{moa}
Sumi T. et al., 2003, ApJ, 591, 204

\bibitem[\protect\citeauthoryear{Sumi et al.}{2006}]{ogle}
Sumi T. et al., 2006, ApJ, 636, 240

\bibitem[\protect\citeauthoryear{Twarog}{1980}]{twa80}
Twarog B., 1980, ApJS, 44, 1

\bibitem[\protect\citeauthoryear{Udalski et al.}{2002}]{uda02}
Udalski A. et al., 2002, AcA, 52, 217

\bibitem[\protect\citeauthoryear{Udalski et al.}{2003}]{uda03}
Udalski A. et al., 2003, AcA, 53, 291

\bibitem[\protect\citeauthoryear{Wood \& Mao}{2005}]{wood05}
Wood A., Mao S., 2005, MNRAS, 362, 945


\end{thebibliography}
\end{document}